\documentclass[
aps,%
12pt,%
final,%
notitlepage,%
oneside,%
onecolumn,%
nobibnotes,%
nofootinbib,%
showpacs,%
superscriptaddress,%
centertags,floatfix]%
{revtex4}

\usepackage{amsmath,amssymb}
\usepackage{bm}

\begin{document}

\title{ON LIMITS OF {\it ab initio} CALCULATIONS  OF PAIRING
GAP IN NUCLEI}

\author{\firstname{E.E.}~\surname{Saperstein}}
\email{saper@mbslab.kiae.ru} \affiliation{Kurchatov Institute,
123182 Moscow, Russia.}
\author{\firstname{M.}~\surname{Baldo}}
\affiliation{INFN, Sezione di Catania, 64 Via S.-Sofia, I-95123
Catania, Italy.}
\author{\firstname{U.}~\surname{Lombardo}}
\affiliation{INFN-LNS and
University of Catania, 44 Via S.-Sofia, I-95125 Catania, Italy.}
\author{\firstname{S.S.}~\surname{Pankratov}}
\affiliation{Kurchatov Institute, 123182 Moscow, Russia; Moscow Institute
of Physics and Technology, 123098 Moscow, Russia.}
\author{\firstname{M.V.}~\surname{Zverev}}
\affiliation{Kurchatov Institute, 123182 Moscow, Russia;\\ Moscow
Institute of Physics and Technology, 123098 Moscow, Russia.}

\begin{abstract}
A brief review of recent  microscopic calculations of nuclear
pairing gap is given. A semi-microscopic model is suggested in which
the {\it ab-initio} effective pairing interaction is supplemented
with a small phenomenological addendum. It involves a parameter
which is universal for all medium and heavy nuclei. Calculations for
several isotopic and isotonic chains of semi-magic nuclei confirm
the relevance of the model.
\end{abstract}

\maketitle

\section{Introduction}

Up to now, there is no consistent microscopic theory of nuclear
matter. The well-known Brueckner theory \cite{R_Sch} was the first
very promising step in this direction but next steps are very
complicated as one deals with the many-body problem without any
small parameter. Why the idea to develop the {\it ab initio} theory
of pairing in finite nuclei is not absolutely unreasonable, although
a finite nucleus is much more complicated system than infinite
nuclear matter? The point is that, for the pairing problem, some
simplifications occur in finite nuclei. They originate from the
surface nature of nuclear pairing \cite{Rep}. If the pairing problem
is formulated in terms of an effective pairing interaction $\Vef$ in
a model space $S_0$, this quantity turns out to be density dependent
\cite{Fay,HFB} with strong dominance of the surface attraction. To
be more definite, let us write down the simple local 2-parameter
ansatz   for $\Vef$ within the Finite Fermi Systems (FFS) theory
\cite{AB}: \beq \Vef({\bf r}_1,{\bf r}_2,{\bf r}_3,{\bf r}_4) =C_0
\left[\gamma^{\rm ex} + (\gamma^{\rm in} - \gamma^{\rm ex} ) \frac
{\rho(r_1)}{\rho(0)}\right] \delta ({\bf r}_1 - {\bf r}_2)\delta
({\bf r}_1 - {\bf r}_3)\delta ({\bf r}_2 - {\bf r}_4).
\label{VefFFS} \eeq Here $C_0=300\;$MeV$ \cdot$ fm$^3$ is the
inverse density of state at the Fermi surface, the standard FFS
theory dimension factor for the effective interaction, and $\rho(r)$
is the density of the kind of nucleons  under consideration. Typical
values of the parameters (e.g. in \cite{Fay}) correspond to the
external constant $\gamma^{\rm ex}$ approximately ten times greater
than the internal one, $\gamma^{\rm in}$. Therefore it seems
reasonable to try to find $\Vef$ starting from the first principles,
as the conditions for the validity of the Brueckner theory at the
surface are much better than inside nuclei. Within the Brueckner
theory, the gap equation coincides with that of the
Bardeen-Cooper-Shrieffer (BCS) theory, as the ladder diagrams
summation typical of the Brueckner theory is made already in the gap
equation itself. In any case, the Brueckner theory is valid
trivially outside the nucleus where all many-body corrections
vanish, and therefore it should correctly reproduce the $\gamma^{\rm
ex}$ parameter. This is not the case for the $\gamma^{\rm in}$
parameter (or, more generally, for the in-side behavior of the
$\Vef({\bf r}_i)$ function). But, since it is  small, one can hope
that even noticeable corrections to $\gamma^{\rm in}$ should not
 significantly change the
 gap $\Delta$ value. Such a logic
has a weak point because of the exponential dependence of the gap on
the interaction strength which is well known in the weak coupling
limit of the BCS theory: \beq \Delta_{\rm F} \, \approx\, 2
\eps_{\rm F} \exp (1/\nu_{\rm F}\Vef )\;, \label{expf}\eeq  where
$\nu_{\rm F}=m^*k_{\rm F}/\pi^2$ and $\eps_{\rm F}=k_{\rm
F}^2/(2m^*)$, $m^*$ being the effective mass. This is the reason why
the knowledge of the in-side behavior of $\Vef$ is important for
accurate evaluation of the gap value, hence corrections to the BCS
theory should be accounted for.

   In the last few years, some progress has been made in the microscopic
theory of nuclear pairing by the Milan group \cite{milan2,milan3}
and Duguet et al. \cite{Dug1,Dug2}. And some contradictions revealed
already at the ``BCS level'', although both the calculations were
made within rather close frameworks. In particular, the same
single-particle spectrum was used for solving the gap equation,
namely, it was calculated within the Skyrme--Hartree--Fock (SHF)
method with the Sly4 force which produces the coordinate dependent
effective mass $m^*(r)$ essentially different from the bare one $m$.
In Ref. \cite{milan3}, the value $\Delta\simeq 1.0\;$MeV was found
for the gap in the nucleus $^{120}$Sn (a traditional benchmark for
the pairing problem) which is noticeably less of the experimental
one, $\Delta_{\rm exp}\simeq 1.3\;$MeV. At the same time, in
\cite{Dug1} the value $\Delta\simeq 1.6\;$MeV was obtained for the
same nucleus which is essentially larger. In the first case, a lack
of the gap value is explained in \cite{milan3} by invoking various
many-body corrections to the BCS approximation, exchange with
low-lying surface vibrations (``phonons'') being the main of them.
Indeed, the latter enlarges the gap value (see, e.g., \cite{milan2}
and \cite{Av_Kam}) making it closer to the experimental value. At
the same time, it is rather difficult to find a mechanism that can
reduce the value of $\Delta$ in Ref. \cite{Dug1}. In Refs.
\cite{Pankr1,Bald1} we have analyzed the reasons of these
contradictions. This point was discussed also in \cite{Dug2}. It
turned out that, in fact, these two calculations differ in the way
they take into account the effective mass. It implies that the gap
$\Delta$  depends not only on the value of the effective mass at the
Fermi surface, as it follows from Eq. (\ref{expf}), but also on the
behavior of the function $m^*(k)$ in a wide momentum range. But this
quantity is not known sufficiently well \cite{Bald1} that makes
rather uncertain the predictions of such calculations. To avoid it,
we suggest a semi-microscopic model for the effective pairing
interaction in which the main ab-initio term of $\Vef$ is
supplemented with a small addendum containing one phenomenological
parameter. Preliminary results of this model were presented in
\cite{Pankr2}.

\section{Outline of the formalism}

The general form of the many-body theory equation for the pairing
gap $\Delta$ reads \cite{AB}:
\beq \Delta =  {\cal U} G G^s \Delta, \label{del} \eeq where ${\cal
U}$ is the $NN$-interaction block irreducible in the two-paricle
channel, and
 $G$ ($G^s$) is the one-particle Green function without (with) pairing effects
 taken into account. A symbolic multiplication, as
usual, denotes the integration over energy and intermediate
coordinates and summation over spin variables as well. When we used
above the term  ``BCS theory'', we meant to replace the block ${\cal
U}$ of irreducible interaction diagrams with the free $NN$-potential
${\cal V}$  in Eq. (\ref{del}) and to use the simple quasiparticle
Green functions for $G$ and $G^s$ (e.g, without phonon corrections
or others). In this case, Eq. (\ref{del}) is greatly simplified and
can be reduced to the form usual for the Bogolyubov method, \beq
\Delta = - {\cal V} \varkappa, \label{delkap} \eeq where
\beq\varkappa=\int \frac {d\eps}{2\pi i}G G^s\Delta
 \label{defkap}\eeq is the anomalous density matrix
which can be expressed explicitly in terms of the Bogolyubov
functions $u$ and $v$,
\beq \varkappa({\bf r}_1,{\bf r}_2) = \sum_i u_i({\bf r}_1) v_i({\bf
r}_2). \label{kapuv} \eeq Summation in (\ref{kapuv}) is carried out
over the complete set of Bogolyubov  functions with eigen energies
$E_i>0$.

In Refs. \cite{milan2,milan3}, the set of Bogolyubov equations,
together with the gap equation  (\ref{del}) with the realistic
Argonne $NN$-interaction v$_{14}$, was solved directly in the basis
\{$\lambda$\} of states restricted to the energy domain up to
$E_{\max}{=}800\;$MeV. In addition, as mentioned above, the SHF
basis with the  SLy4 force was used with the coordinate dependent
effective mass $m^*(r)$, which is considerably smaller than the bare
mass $m$. The main difficulty of the direct method to solve the
nuclear pairing problem comes from rather slow convergence of the
sums over intermediate states $\lambda$ in the gap equation because
of the short-range of the free $NN$-force. Evidently, this is the
reason why the authors of  \cite{milan2,milan3} limited the
calculations only to one nucleus $^{120}$Sn. To avoid the slow
convergence problem, the authors of \cite{Dug1,Dug2} used the
super-soft ``low-k'' force $V_{\rm low-k}$ \cite{Kuo} which is
defined in such a way that it describes correctly the
 $NN$-scattering phase shifts at momenta  $k{<}\Lambda$, where $\Lambda$
 is a parameter which is not bigger than the one corresponding to
 the limiting energy  $E_{\rm
lim} \simeq 300\;$MeV, for smaller energy values the phase shifts
being reproduced accurately. As the force $V_{\rm low-k}$ vanishes
rapidly for $k{>}\Lambda$, one can limit the energy up to $E_{\max}
{\simeq} 300\;$MeV in the gap equation  (\ref{delkap}). This made it
possible to calculate in  \cite{Dug1} neutron and proton pairing
gaps for a lot of nuclei. Usually the low-k force is found starting
from some realistic  $NN$-potential ${\cal V}$ with the help of the
Renormaliation Group method, and the result doesn't practically
depend on the particular choice  of  ${\cal V}$ \cite{Kuo}. In
addition, in \cite{Dug1} $V_{\rm low-k}$ was found starting from the
Argonne potential v$_{18}$, which is different only a little from
the one used in \cite{milan3}, v$_{14}$. Thus, indeed, the schemes
of solving the BCS gap equation in  \cite{Dug1} and \cite{milan3}
were very similar.

To overcome the slow convergence problem in the gap equation for
finite systems, we used a two-step renormalization method.
 In this approach, we split the complete Hilbert
space of the pairing problem  $S$ to the model subspace $S_0$,
including the single-particle states with energies less than a fixed
value of $E_0$, and the subsidiary one, $S'$. The gap equation is
solved in the model space: \beq \Delta = \Vef G G^s \Delta|_{S_0},
\label{del0} \eeq with the effective pairing interaction  $\Vef$
instead of the block ${\cal U}$ in the original gap equation
(\ref{del}). It obeys the Bethe--Goldstone type equation in the
subsidiary space,  \beq \Vef = {\cal U} + {\cal U} G G \Vef|_{S'}.
\label{Vef} \eeq In this equation, the pairing effects could be
neglected provided the model space is sufficiently large. That is
why we replaced the Green function $G^s$ for the superfluid system
with its counterpart  $G$ for the normal system. In the BCS
approximation, the block  ${\cal U}$ in (\ref{Vef}) should be
replaced by  ${\cal V}$. To solve equation (\ref{Vef}) in
non-homogeneous  systems, we have found a new form of the local
approximation, the Local Potential Approximation (LPA). Originally
it was developed for semi-infinite nuclear matter \cite{Bald0}, then
for the slab of nuclear matter (see review articles \cite{Rep,ST})
and finally, for finite nuclei \cite{Pankr1,Bald1}. It turned out
that, with very high accuracy, at each value of the c.m. coordinate
${\bf R}$, in Eq. (\ref{Vef}) the formulae  of the infinite system
embedded into the constant potential well $U=U({\bf R})$ (it
explains the term LPA) can be used. This simplifies equation for
$\Vef$ significantly, in comparison with the initial equation for
$\Delta$. As the result, the subspace $S'$ can be chosen as large as
necessary. From the comparison of the direct solution of Eq.
(\ref{Vef}) in the slab with the LPA one, it was shown  that the LPA
has high accuracy, even in the surface region, for sufficiently
large model space, $E_0$ (${\simeq} 20{\div} 30\;$MeV). For finite
nuclei (the same $^{120}$Sn), validity of LPA was checked also
\cite{Pankr1,Bald1}. In this case, the boundary energy should be
made larger up to $E_0{=}40\;$MeV. In this article, we use the LPA
with this value of $E_0$ for systematic calculations of the gap in
spherical nuclei. For ${\cal V}$, we use  just as in \cite{Bald1},
the Argonne potential  v$_{18}$.

Let us note that the use of the low-k force $V_{\rm low-k}$ could be
also interpreted in terms of the two-step renormalization scheme of
solving the gap equation (\ref{del}), with  $E_0 {\simeq} 300\;$MeV
and with free nucleon Green functions  $G$ in (\ref{Vef}) (i.e.
$U(R)=0$). Then, (with ${\cal U}{\to} {\cal V}$) one obtains
$\Vef{\to} V_{\rm low-k}$ (see \cite{Kuo-Br} where the usual
renormalization scheme, similar to ours, is used to find $V_{\rm
low-k}$ instead of the Renormalization Group equation). Now, the
comparison of the direct solution of the gap equation (\ref{del})
(or (\ref{delkap}))  in Ref. \cite{milan3}  with the Argonne
$NN$-potential ${\cal V}$ and of ``renormalized'' equation
(\ref{del0}) with $\Vef = V_{\rm low-k}$ shows that the difference
appears because, in the subsidiary subspace $S'$, the effective mass
$m^*{\neq} m$ is used in the first case and $m^*=m$, in the second
one. Thus, the result for the gap depends not only on the value of
the effective mass at the Fermi surface, but also on the behavior of
the function $m^*(k)$ in a wide momentum range. This dependence was
demonstrated explicitly in \cite{Pankr1,Bald1}. The use of the SHF
effective mass corresponding to the SLy4 force, or to any other
version of the Skyrme force, could hardly be approved. Indeed, these
effective forces were introduced and fitted to describe
systematically nuclear masses and radii. As a rule, the description
of the single-particle spectrum nearby the Fermi surface with Skyrme
forces is rather poor, and furthermore it is difficult to expect
that they will reproduce it correctly at those high momenta that are
involved in the gap equation (\ref{del}). This point makes it
problematic the problem of finding the pairing gap from the first
principles completely. The situation is even more dramatic because
the many-body theory equation (\ref{del}) contains, in addition to
the ``$k$-mass'' of the SHF method, the ``$E$-mass'' (inverse
$Z$-factor) \cite{bg1,bg2,LSZ}, which also is not sufficiently well
known even in nuclear matter \cite{Bald1}. The  corrections to the
BCS version of Eq. (\ref{del}) include also the difference of the
block ${\cal U}$ from the polential ${\cal V}$, mainly due to the
so-called induced interaction. The attempt in \cite{milan3} to find
it in terms of the same SLy4 force as the nuclear mean field looks
questionable. Indeed, this force was fitted to the nuclear
characteristics which depend mainly on those Skyrme parameters
determining the scalar Landau--Migdal (LM) amplitudes $f,f'$. As to
the spin amplitudes $g,g'$, they remain practically undetermined in
the SHF method. At the same time, the contribution of the spin
channel to the induced interaction is not less than of the scalar
one \cite{milan3}. Parameters $g,g'$ are well known from the
calculations of nuclear magnetic moments within the Finite Fermi
Systems (FFS) theory \cite{BST}, but, just as the Skyrme parameters,
at the Fermi surface only. But the states distant from the Fermi
surface are important to calculate the induced interaction. At last,
let us imagine to get from some phenomenology the functions $m^*(k),
Z(k)$ and all the LM amplitudes far from the Fermi surface. Even in
this case, the use of so many phenomenological ingredients devalues
significantly the {\it ab initio} starting point, i.e. the free
$NN$-potential ${\cal V}$ in the pairing gap calculation.

Instead, we suggest to introduce in the effective pairing
interaction a small phenomenological addendum which embodies, of
course approximately, all the corrections to the BCS scheme
discussed above. The simplest ansatz for it is similar to Eq.
(\ref{VefFFS}) and reads: \beq {\cal V}_{\rm eff} = V^{0}_{\rm eff}
+ \gamma C_0 \frac {\rho(r_1)}{\bar{\rho}(0)}\delta ({\bf r}_1 -
{\bf r}_2)\delta ({\bf r}_1 - {\bf r}_3)\delta ({\bf r}_2 - {\bf
r}_4). \label{Vef1} \eeq Here   $\rho(r)$ is the density of nucleons
of the kind under consideration, and $\gamma$ is a dimensionless
phenomenological parameter. To avoid any influence of the shell
fluctuations in the value of ${\rho}(0)$, ${\bar{\rho}(0)}$ was
averaged over the interval of $r{<}2\;$fm. The first, {\it ab
initio}, term in the r.h.s. of  Eq. (\ref{Vef1}) is the solution of
Eq. (\ref{Vef}) with ${\cal U} {=} {\cal V}$ in the framework of the
LPA method described above, with $m^*{=}m$ in the subspace $S'$.
Then, the gap equation (\ref{del0}) in the model space is solved
with the self-consistent basis found within the Generalized Energy
Density Functional (GEDF) method \cite{Fay} with the functional DF3
where the identity $m^*{=}m$ is assumed. The latter is of principal
importance for our approach. First, it makes the results less
model-dependent, all effects of $m^*\neq m$ in both  model and
subsidiary subspaces being attributed to the in-medium corrections
beyond the pure BCS approximation. Second, single-particle spectra
of the GEDF method \cite{Fay} are, as a rule, in better agreement
with the experimental ones  than those of the popular versions of
the SHF method \cite{TS}. The quality of the single-particle
spectrum nearby the Fermi surface is very important for obtaining
the correct value of the gap found from Eq. (\ref{delkap}).

\section{On the procedure to find the ``experimental'' gap }

The gap $\Delta$ is not an observable quantity which can be
extracted from experimental data directly. Usually, this quantity,
 $\Delta_{\rm exp}$, is found in terms of
 mass values $M$ of neighboring nuclei via 3-term formulae,
 \beq 2\Delta^+_{\rm exp}(A)= \delta_2M^+ \equiv
 2M(A+1)-M(A+2)- M(A),\label{dexp1}\eeq or
\beq 2\Delta^-_{\rm exp}(A)= \delta_2M^- \equiv 2M(A-1)-
M(A-2)-M(A).\label{dexp2}\eeq The 5-term expression is usually
considered  more accurate, being a half-sum of them, \beq
\Delta_{\rm exp}(A)=  \overline{\delta_2 M}/2 \equiv (\delta_2M^+ +
\delta_2M^-)/2.\label{dexp}\eeq These simple recipes were used, in
particular, in \cite{milan2,milan3,Dug1,Dug2}. However, they
originate from the simplest model $\Delta=const$, and  the accuracy
of such prescription  is not obvious {\it a priori}. To clarify this
point we made a calculation which could be considered as a
``theoretical experiment''. We used the GEDF method \cite{Fay} with
the functional DF3 which reproduces the mass differences of Eqs.
(\ref{dexp1}),(\ref{dexp2}) type sufficiently well. First, we
calculated the right side of Eq. (\ref{dexp}) directly, and second,
the theoretical gap value. For the latter, we use the ``Fermi
average'' combination, \beq
\DF{=}\sum_{\lambda}{(2j{+}1)\Delta_{\lambda
\lambda}}/\sum_{\lambda}(2j{+}1), \label{DelF}\eeq where the
summation is carried out over the states $\lambda$ in the interval
of $|\eps_{\lambda}{-}\mu|{<}3\;$MeV. A similar recipe  was used,
e.g., in \cite{milan3}. The comparison of these two quantities  is
given in fig. 1 for the lead isotopes and in fig. 2 for the tin
isotopes. We see that for the main part of nuclei under
consideration the difference between values in two neighboring
columns is within 0.1 MeV. However, there is several cases where it
is of the order (or even exceeds) 0.2 MeV. Leaving aside detailed
analysis of these ``bad'' cases we are forced to put a limit of
$\simeq 0.1 - 0.2\;$MeV in the accuracy of the experimental gap
determined from Eq. (\ref{dexp}).

\begin{figure}
\setcaptionmargin{5mm}
\onelinecaptionstrue
\includegraphics[height=80mm,width=100mm]{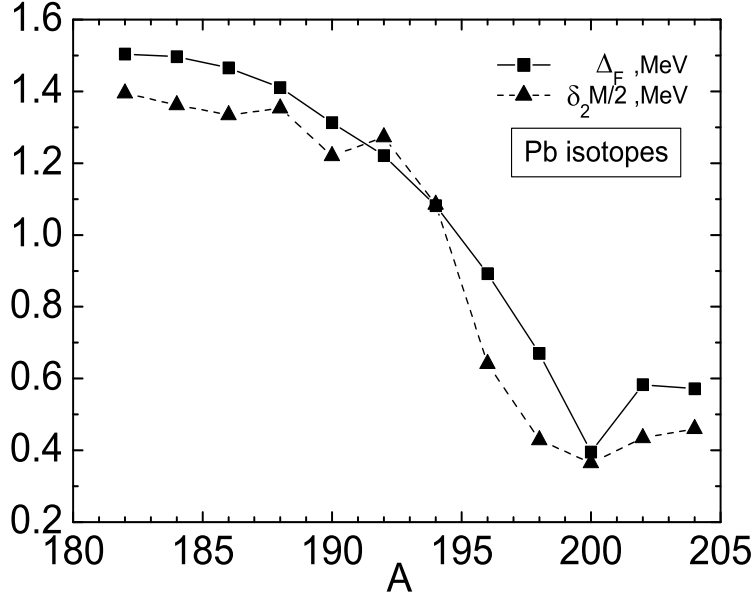}%
\captionstyle{normal} \caption{ Comparison of the theoretical mass
differences $\overline{\delta_2 M}/2 $ with average gap values
$\Delta_{\rm F}$ for Pb isotopes}
\end{figure}

\begin{figure}
\setcaptionmargin{5mm}
\onelinecaptionstrue
\includegraphics[height=80mm,width=100mm]{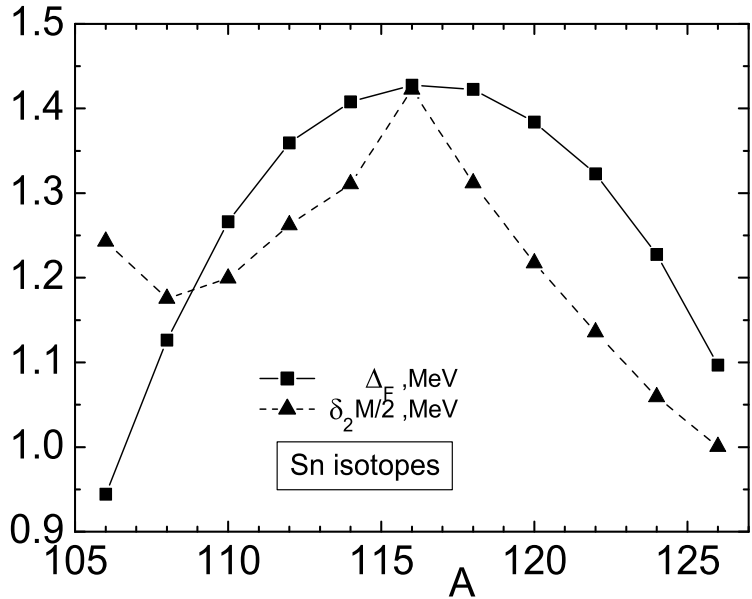}%
\captionstyle{normal} \caption{The same as in fig. 1, but for Sn
isotopes }
\end{figure}

\section{Calculation results}
As it was discussed above, for the model space we used the GEDF
method by Fayans et. al. with the DF3 functional \cite{Fay}. The
model space was extended up to the energy $E_0=40\;$MeV, the
subsidiary one up to $E_{\rm max}=1000\;$MeV. The spherical box of
the radius $R=16\;$fm was used, with the grid step $h=0.05\;$fm. The
numerical stability of the results was checked by increasing the
parameters up to $E_0=60\;$MeV, $E_{\rm max}=1200\;$MeV and
$R=24\;$fm, and we found for the gap value a numerical accuracy of
0.01 MeV.

We calculated the neutron gap for 25 semi-magic isotopes of the
lead, tin and calcium chains and the proton gap in 9 nuclei, also
semi-magic, isotones of the $N=82$ chain. The formulae above
correspond to so-called ``developed pairing'' approximation
\cite{AB}, i.e. imposing the equality of the $\Delta^+$ and
$\Delta^-$ operators. Therefore we limit ourselves to nuclei having,
as a minimum, four particles (holes) above (below) the magic core.
Therefore, the only isotope $^{44}$Ca was considered in the calcium
chain.

\begin{table}[t]
\setcaptionmargin{0mm} \onelinecaptionsfalse
\captionstyle{flushleft} \caption{ Neutron gap $\Delta^n_{\rm F}$
(MeV) in semi-magic nuclei.}

\begin{tabular}{c|c|c}
\hline \hspace*{0.4ex} nucleus \hspace*{0.4ex} & \hspace*{11.5ex} $\Delta^n_{\rm F}$  \hspace*{11.5ex} & \hspace*{0.7ex} $\Delta_{\rm exp}$\hspace*{0.7ex} \\
\hline
\end{tabular}

\begin{tabular}{c|c|c|c|c}
\hline \hspace*{9.7ex} & \hspace*{1.5ex} $\gamma$=0 \hspace*{1.5ex}
& \hspace*{1.5ex} 0.06 \hspace*{1.5ex} & \hspace*{1.5ex} 0.08
\hspace*{1.5ex} &
\hspace*{3ex}  \hspace*{3ex} \\

\hline

$^{182}$Pb  & 1.79 & 1.33 & 1.20 & 1.30\\

$^{184}$Pb  & 1.79 & 1.33 & 1.20 & 1.34\\

$^{186}$Pb  & 1.78 & 1.32 & 1.19 & 1.30\\

$^{188}$Pb  & 1.76 & 1.31 & 1.17 & 1.25\\

$^{190}$Pb  & 1.73 & 1.29 & 1.16 & 1.24\\

$^{192}$Pb  & 1.68 & 1.22 & 1.09 & 1.21\\

$^{194}$Pb  & 1.62 & 1.16 & 1.03 & 1.13\\

$^{196}$Pb  & 1.53 & 1.09 & 0.96 & 1.01\\

$^{198}$Pb  & 1.43 & 1.00 & 0.87 & 0.94\\

$^{200}$Pb  & 1.31 & 0.90 & 0.80 & 0.87\\

$^{202}$Pb  & 1.16 & 0.79 & 0.69 & 0.78\\

$^{204}$Pb  & 0.95 & 0.64 & 0.56 & 0.71\\

\hline

$^{106}$Sn  & 1.35 & 0.95 & 0.83 & 1.20\\

$^{108}$Sn  & 1.52 & 1.13 & 1.01 & 1.23\\

$^{110}$Sn  & 1.65 & 1.26 & 1.14 & 1.30\\

$^{112}$Sn  & 1.74 & 1.34 & 1.23 & 1.29\\

$^{114}$Sn  & 1.80 & 1.40 & 1.28 & 1.14\\

$^{116}$Sn  & 1.82 & 1.43 & 1.31 & 1.10\\

$^{118}$Sn  & 1.83 & 1.44 & 1.32 & 1.25\\

$^{120}$Sn  & 1.80 & 1.42 & 1.31 & 1.32\\

$^{122}$Sn  & 1.74 & 1.38 & 1.28 & 1.30\\

$^{124}$Sn  & 1.65 & 1.30 & 1.21 & 1.25\\

$^{126}$Sn  & 1.51 & 1.19 & 1.10 & 1.20\\

$^{128}$Sn  & 1.31 & 1.02 & 0.94 & 1.16\\

\hline

$^{44}$Ca   & 1.83 & 1.50 & 1.41 & 1.54\\

\hline
\end{tabular}\label{tab_N}
\end{table}


\begin{figure}
\setcaptionmargin{5mm}
\onelinecaptionstrue
\includegraphics[height=80mm,width=100mm]{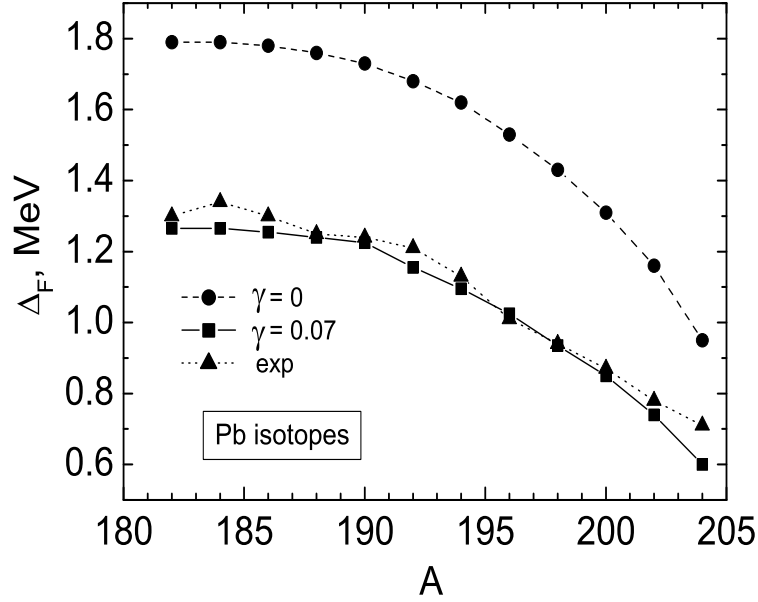}%
\captionstyle{normal} \caption{Neutron gap in Pb isotopes}
\end{figure}

\begin{figure}
\setcaptionmargin{5mm}
\onelinecaptionstrue
\includegraphics[height=80mm,width=100mm]{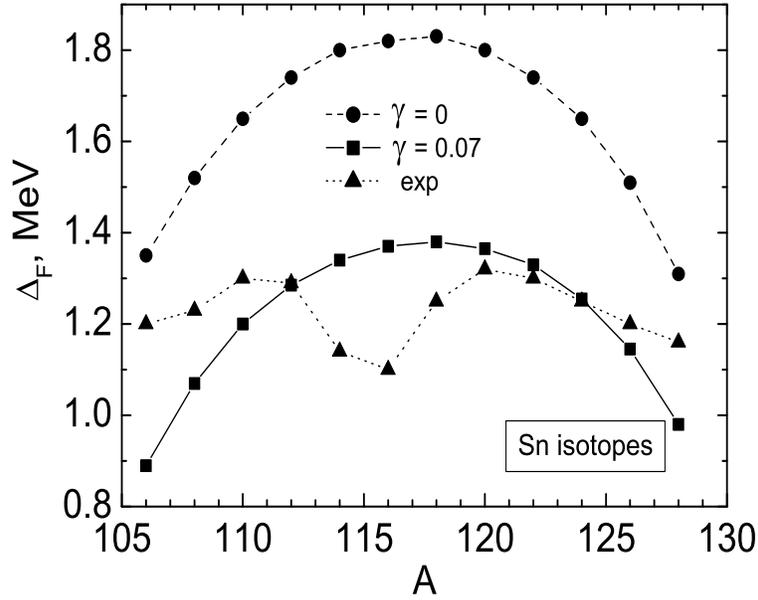}%
\captionstyle{normal} \caption{Neutron gap in Sn isotopes}
\end{figure}

Let us begin with the neutron pairing. The results are presented in
table 1 and figs. 3,4. The Fermi average gap values, Eq.
(\ref{DelF}), found for different values of the parameter $\gamma$
in Eq. (\ref{Vef1}). We see that the gap values with the  ``ab
initio'' interaction ($\gamma{=}0$) are greater by 30 -- 40\%  than
the experimental ones. This difference exceeds significantly the
accuracy of $\simeq 0.1 - 0.2\;$MeV for the gap value which we could
expect in accordance with discussion of the previous section. As it
can be seen, with few exceptions, it is obtained for $\gamma = 0.06
- 0.08$. For the ``optimal'' value of $\gamma = 0.07$ (the results
are exactly half-sums of the values in the third and forth columns),
the theoretical error exceeds this limit only in $^{106}$Sn and
$^{116}$Sn. Evidently, it is caused by the fact that the DF3
functional provides an incorrect reproduction  of the
``intruder''-state $1h_{11/2}$, which plays an essential role in the
gap equation (\ref{del0}) for these nuclei. Fig. 3 and fig. 4 are
drawn just to illustrate the optimal value of $\gamma$. To show that
the phenomenological addendum to the effective pairing interaction
in (\ref{Vef1}) is indeed rather small for $\gamma = 0.07$, we
displayed in fig. 5 the localized ``Fermi average'' of the effective
interaction. In the mixed coordinate-momentum representation, it is
defined as follows: ${\cal V}_{\rm eff}({\bf k}_1,{\bf k}_2,{\bf
r}_1,{\bf r}_2)\to {\cal V}^{\rm F}_{\rm eff}(R=r_1) \delta({\bf
r}_1-{\bf r}_2) \delta({\bf r}_1-{\bf r}_3) \delta({\bf r}_2-{\bf
r}_4)$, where \beq {\cal V}^{\rm F}_{\rm eff}(R)= \int d^3t {\cal
V}_{\rm eff}(k_1=k_2=k_{\rm F}(R),{\bf R}-{\bf t}/2,{\bf R}+{\bf
t}/2),\eeq with $k_{\rm F}(R)=\sqrt{2m(\mu-U(R))}$, provided
$\mu-U(R)\ge 0$, and $k_{\rm F}(R)=0$ otherwise. Here $\mu$ and
$U(R)$ are the chemical potential and the potential well of the kind
of nucleons  under consideration. A similar quantity was considered
before in the slab system to visualize the effective interaction
properties \cite{Rep,EPI}. At a glance, the difference between the
interaction strengths for $\gamma{=}0$ and $\gamma{=}0.07$ is
negligible, but it produces noticeable effects in the gap due to the
exponential behavior in Eq. (\ref{expf}).

\begin{figure}
\setcaptionmargin{5mm}
\onelinecaptionstrue
\includegraphics[height=80mm,width=100mm]{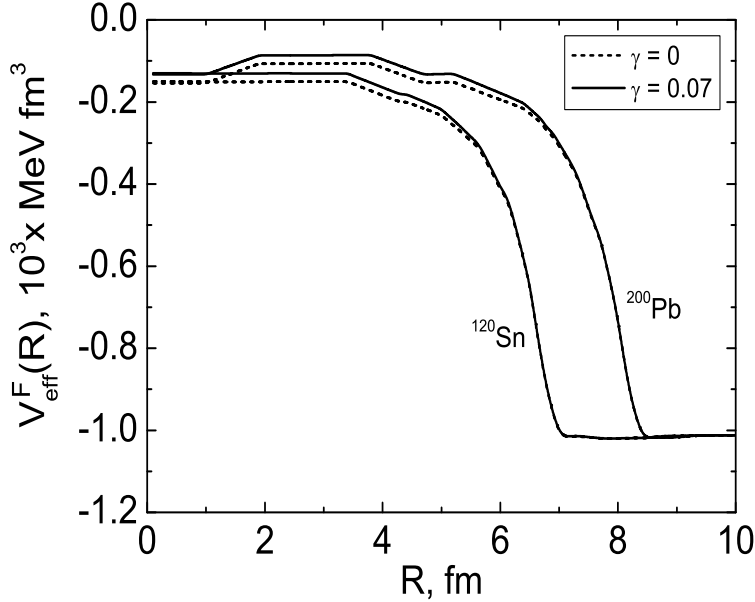}%
\captionstyle{normal} \caption{Fermi-average effective pairing
interaction}
\end{figure}

Let us now turn to protons. In this case, the Coulomb potential
${\cal V}_{\rm C}$ must be added to the expression (\ref{Vef1}),
\beq {\cal V}_{\rm eff}^p={\cal V}_{\rm eff}^n+{\cal V}_{\rm C}.\eeq
Again this addendum is small and again it turned out to be important
for the gap equation due to the the enhancement discussed above. In
particular, this was demonstrated in previous calculations
\cite{Dug1,Dug2}. The estimates show that the Coulomb potential
could be taken in the bare form. Indeed, in the momentum space one
has ${\cal V}_{\rm C}=e^2/q^2$ with a strong maximum at small $q$
values provided they persist in the matrix elements
$<\lambda_1\lambda_2| {\cal V}_{\rm C}|\lambda_3\lambda_4>$, with
obvious notation. In the gap equation, the diagonal elements with
$\lambda_1=\lambda_2 =\lambda_3=\lambda_4$ are of primary importance
for which the region around $q\simeq 0$ in the integral dominates.
But at small $q$ the Coulomb potential ${\cal V}_{\rm C}(q)$ is not
modified due to the Ward identity. In non-diagonal matrix elements
the  contribution of $q\simeq k_{\rm F}$ dominates and ${\cal
V}_{\rm C}(q)$ could be modified, but in this case the contribution
of ${\cal V}_{\rm C}$ is very small and can be neglected.

\begin{table}[t]
\setcaptionmargin{0mm} \onelinecaptionsfalse
\captionstyle{flushleft} \caption{ Proton gap $\Delta^p_{\rm F}$
(MeV) for the isotone gap $N=82$.}
\bigskip

\begin{tabular}{c|c|c}
\hline \hspace*{0.4ex} nucleus \hspace*{0.4ex} & \hspace*{15.5ex} $\Delta^p_{\rm F}$  \hspace*{15.3ex} & \hspace*{0.7ex} $\Delta_{\rm exp}$\hspace*{0.7ex} \\
\hline
\end{tabular}

\begin{tabular}{c|c|c}
\hline \hspace*{7.9ex} ${\cal V}_{\rm eff}^p{=}{\cal V}_{\rm eff}^0$
& \hspace*{5.7ex}
${\cal V}_{\rm eff}^p{=}{\cal V}_{\rm eff} + {\cal V}_{\rm C} $ \hspace*{6.ex} & \hspace*{7.ex} \\
\hline
\end{tabular}

\begin{tabular}{c|c|c|c|c|c}
\hline \hspace*{9.7ex} & \hspace*{3ex}  \hspace*{3ex} &
\hspace*{1.5ex} $\gamma$=0 \hspace*{1.5ex} & \hspace*{1.5ex} 0.06
\hspace*{1.5ex} & \hspace*{1.5ex} 0.08 \hspace*{1.5ex} &
\hspace*{3ex}  \hspace*{3ex} \\

\hline

$^{136}$Xe  & 1.65 & 1.19 & 0.87 & 0.78 & 0.75\\

$^{138}$Ba  & 1.80 & 1.33 & 0.98 & 0.88 & 0.87\\

$^{140}$Ce  & 1.90 & 1.42 & 1.03 & 0.92 & 0.97\\

$^{142}$Nd  & 1.99 & 1.48 & 1.06 & 0.94 & 1.00\\

$^{144}$Sm  & 2.01 & 1.49 & 1.05 & 0.91 & 1.02\\

$^{146}$Gd  & 2.02 & 1.50 & 1.05 & 0.91 & 1.13\\

$^{148}$Dy  & 2.01 & 1.50 & 1.06 & 0.93 & 1.19\\

$^{150}$Er  & 1.98 & 1.48 & 1.07 & 0.94 & 1.22\\

$^{152}$Yb  & 1.92 & 1.44 & 1.05 & 0.93 & 1.29\\

\hline

\end{tabular}\label{tab_N}
\end{table}


\begin{figure}
\setcaptionmargin{5mm}
\onelinecaptionstrue
\includegraphics[height=80mm,width=100mm]{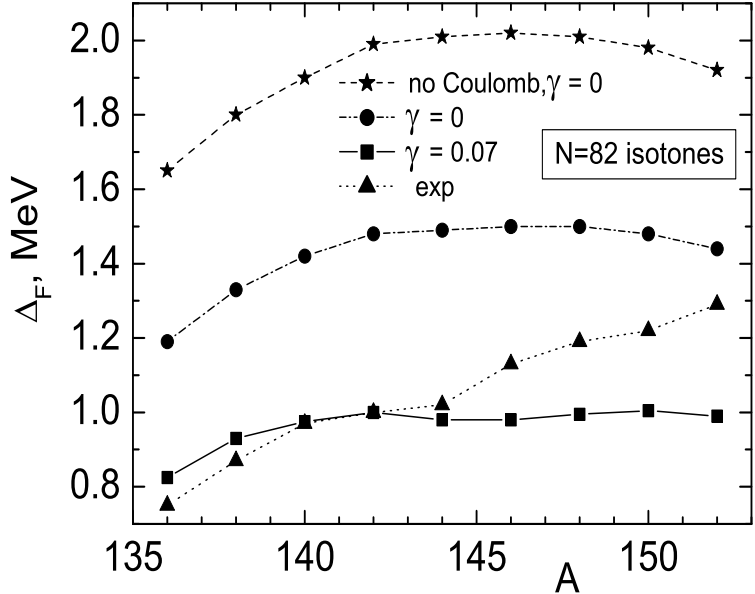}%
\captionstyle{normal} \caption{Proton gap in $N=82$ isotones}
\end{figure}

The  results for the isotone chain $N{=}82$ are given in Table 2 and
displayed in fig. 6. To demonstrate the effect of the Coulomb
interaction, we show the results with the interaction  ${\cal
V}_{\rm eff}^p={\cal V}_{\rm eff}^0$ which difference from the
corresponding value in column 3 gives exactly the Coulomb effect in
the gap. Indeed, it is rather big (about 0.5 MeV), in qualitative
agreement with \cite{Dug1}. Again at $\gamma {=}0.07$ the agreement
is almost perfect for the most part of nuclei, and only for the two
heaviest isotones the disagreement exceeds 0.2 MeV. In this case,
the possible reason lies in the proximity to the phase transition to
the deformed state (at $A\simeq 150$). Average difference between
the theoretical and experimental gap values for 34 nuclei considered
is equal to $\sqrt{\overline{(\delta \Delta)^2}}{\simeq}0.13\;$MeV.
As it follows from the analysis in Sect. 3, this value is within the
accuracy of the experimental values of the gap defined with the
relation (\ref{dexp}).

\section{Conclusions}
We suggest a simple semi-microscopic model (\ref{Vef1}) for the
effective pairing interaction containing one phenomenological
parameter which takes into account approximately various corrections
to the pure BCS theory. This model reproduces rather well
experimental values of the neutron and proton gaps in semi-magic
nuclei. The overall  agreement ($\sqrt{\overline{(\delta
\Delta)^2}}{\simeq}0.13\;$MeV) is better than that obtained in
\cite{Dug1}, where the authors did not introduce free parameters
explicitly but they made it implicitly by using a  specific
k-dependence of the effective mass.

The ansatz of Eq. (\ref{Vef1}) possesses an obvious drawback. The
phenomenological GEDF pairing interaction of \cite{Fay} contains the
surface term  (${\propto}(d\rho /dr)^2$) that plays an essential
role for the description of the odd-even effect (staggering) in
nuclear radii. It originates mainly from the exchange by surface
phonons which was explicitly taken into account in
\cite{milan2,milan3}. The addition of such a term in Eq.
(\ref{Vef1}) is associated with introducing a new parameter, and at
the first stage we preferred  to avoid it. A more consistent scheme
should, evidently, include the explicit consideration of the
low-lying phonons, as e.g. in \cite{milan2}, but  taking into
account the so-called tadpole diagrams \cite{Kam_S}. In this case,
the phenomenological constant $\gamma$, of course, will change.

The authors thank G. Colo, T. Duguet and V.A. Khodel for valuable
discussions. This research was partially supported by the  joint
Grants of RFBR and DFG, Germany, No. 09-02-91352-NNIO\_а, 436 RUS
113/994/0-1(R), by the Grants NSh-7235.2010.2  and 2.1.1/4540 of the
Russian Ministry for Science and Education, and by the RFBR grants
09-02-01284-a, 09-02-12168-ofi\_m.


\begin{thebibliography}{16}
\bibitem{R_Sch} P. Ring, P. Schuck, {\it The nuclear many-body
problem} (Springer, Berlin, 1980).

\bibitem{Rep} Baldo M, Lombardo U, Saperstein E E, and Zverev M V
2004  {\it Phys. Rep.} {\bf 391} 261

\bibitem{Fay}
S.A. Fayans, S.V. Tolokonnikov, E.L. Trykov, and D. Zawischa, Nucl.
Phys. A {\bf 676}, 49 (2000).

\bibitem{HFB} S. Goriely, N. Chamel, and J. M. Pearson, Phys. Rev. Lett.
 {\bf 102}, 152503 (2009).

\bibitem{AB} Migdal A B {\it Theory of finite Fermi systems and applications to
atomic nuclei} (Wiley, New York, 1967).

\bibitem{milan2}
 F.~Barranco, R.A. Broglia, G. Colo, {\it et al}., Eur. Phys. J. A {\bf 21}, 57
(2004).

\bibitem{milan3}
 A.~Pastore, F.~Barranco, R.A. Broglia, and E. Vigezzi,  Phys. Rev. C {\bf 78},
 024315 (2008).

\bibitem{Dug1}
T. Duguet and T. Lesinski, Eur. Phys. J. Special Topics {\bf 156},
207 (2008).

\bibitem{Dug2}
K.~Hebeler, T.~Duguet, T.~Lesinski, and A. Schwenk, Phys. Rev. C
{\bf 80}, 044321  (2009).

\bibitem{Av_Kam} A.V. Avdeenkov, S.P. Kamerdzhiev, JETP Lett. {\bf 69}, 669 (1999).

\bibitem{Pankr1}
S.S. Pankratov, M.~Baldo,  M.V. Zverev, U. Lombardo,
E.E.~Saperstein, S.V. Tolokonnikov, JETP Lett., {\bf 90}, 612
(2009).

\bibitem{Bald1}
M. Baldo, U. Lombardo,  S.S. Pankratov, E.E. Saperstein.  J. Phys.
G: Nucl. Phys., 37, 064016 (2010).

\bibitem{Pankr2}
S.S. Pankratov, M.~Baldo,  M.V. Zverev, U. Lombardo,
E.E.~Saperstein, JETP Lett., {\bf 92}, 92 (2010).

\bibitem{Kuo}
S.K. Bogner, T.T.S. Kuo, and A. Schwenk, Phys. Rep. {\bf 386}, 1
(2003).

\bibitem{Bald0}
M. Baldo, U. Lombardo,  E.E. Saperstein, M.V. Zverev, Nucl. Phys. A
{\bf 628} (1998) 503..

\bibitem{ST}
E.E.~Saperstein, S.S. Pankratov, M.V.~Zverev, M.~Baldo,
U.~Lombardo, Phys. At. Nucl., {\bf 72}, 1059 (2009).

\bibitem{Kuo-Br} L.-W. Siu, J.W. Holt, T.T.S. Kuo, and G.E. Brown,
Phys. Rev. C {\bf 79}, 054004 (2009).

\bibitem{bg1}
M. Baldo and  A. Grasso, Phys. Lett. B {\bf 485}, 115 (2000).

\bibitem{bg2}
M. Baldo and  A. Grasso, Phys. At. Nucl. {\bf 64} 611 (2001).

\bibitem{LSZ}
U. Lombardo, P. Schuck, and W. Zuo,
 Phys. Rev. C {\bf 64} 021301(R)  (2001).

\bibitem{BST} V.N. Borzov, E.E. Saperstein, S.V. Tolokonnikov, Phys.
At. Nucl. {\bf 71}, 493 (2008).

\bibitem{KhS}  V.A. Khodel and  E.E. Saperstein, Phys. Rep. {\bf
92},  183 (1982 ).

\bibitem{Kam_S} S. Kamerdzhiev and  E.E. Saperstein, Eur. Phys. J. A {\bf 37},
333 (2008).

\bibitem{TS} S.V. Tolokonnikov and  E.E. Saperstein, Phys.
At. Nucl. {\bf 73}, 000 (2010).

\bibitem{EPI} S.S. Pankratov, M. Baldo, U. Lombardo, E.E. Saperstein, and M.V. Zverev,
 Phys. At. Nucl., {\bf 70}, 688 (2007).

\end{thebibliography}
\end{document}